\documentstyle[preprint,prb,aps]{revtex}
\begin{document}
\draft
\title{Histogram Monte Carlo study of multicritical behavior in the
    hexagonal easy-axis Heisenberg antiferromagnet}
\author{A. Mailhot,\cite{note} M.L. Plumer, and A. Caill\'e}
\address{ Centre de Recherche en Physique du Solide et D\'epartement de
Physique}
\address{Universit\'e de Sherbrooke, Sherbrooke, Qu\'ebec, Canada J1K 2R1}
\date{March 1993}
\maketitle
\begin{abstract}
The results of a detailed histogram Monte-Carlo study of critical-fluctuation
effects on the magnetic-field temperature phase diagram associated with the
hexagonal Heisenberg antiferromagnet with weak axial anisotropy are reported.
The multiphase point where three lines of continuous transitions merge at the
spin-flop boundary exhibits a structure consistent with scaling theory but
without the usual umbilicus as found in the case of a bicritical point.
\end{abstract}
:%============================================================================
% BODY OF PAPER

The frustration of antiferromagnetically coupled sites in a triangular array
gives rise to noncolinear spin order and magnetic-field temperature phase
diagrams which exhibit a rich variety of structure.\cite{plumA,plumB}
In the case of a hexagonal lattice with weak ($c$-axis) axial anisotropy,
a novel type of multicritical point occurs at which three lines of
continuous transitions merge at the spin-flop and paramagnetic phase
boundaries.  Although the full phase diagram has been observed only in the
quasi-one-dimensional hexagonal antiferromagnet $CsNiCl_3$, \cite{john,poir}
similar behavior is expected in\cite{plumA} $RbNiCl_3$, $CsMnI_3$,
\cite{kad} and $CsNiBr_3$.\cite{maeg,chiba}  Gross features of
the experimental results for $CsNiCl_3$ have been reproduced by the analysis
of a phenomenological Landau-type free energy constructed from symmetry
arguments.\cite{plumC}  In zero field, two continuous transitions
occur, identified by the components of the spin vector {\bf M}.\cite{zhu}
As the temperature is lowered from the paramagnetic phase 1,
a linear phase 2 with $M_z$ (where ${\bf z}\|{\bf c}$) is stabilized
at $T_{N1}$, and at $T_{N2}<T_{N1}$ an elliptically polarized phase 3
with $M_z$ and $M_x$ occurs.  In a magnetic field applied along the $c$ axis,
a first-order spin-flop transition to a helically polarized phase 4
with $M_x$ and $M_y$ is found.  Each of the ordered states is characterized
by a period-three modulation in the basal plane and a period-two structure
along the c axis.  (The helical phase 4 is similar to the well-known
$120^\circ$ spin struture of triangular antiferromagnets).  Scaling analysis
of the multicritical point where phase 1, 2, 3, and 4 meet suggests that
the behavior of all the three lines of continuous transitions are governed
by the same crossover exponent $\phi$.\cite{kawaA}  It is the purpose of
the present work to examine in detail this behavior by means of
accurate histogram Monte Carlo simulations of the anisotropic Heisenberg
antiferromagnet.

The easy-axis Hamiltonian studied here is given by \cite{mailA}
\begin{equation}
{\cal H}~=~J_{\|} \sum_{<ij>} {\bf S}_i \cdot {\bf S}_j
\label{eq1}
\end{equation}
where $J_\|>0$ and $J_{\bot}>0$ are the nearest-neighbor antiferromagnetic
exchange interactions along the $c$ axis and in the basal plane, respectively,
$D<0$ is the single-ion anisotropy, and $H$ is magnetic field applied along the
$c$ axis.  Although $J_\|>>J_\bot$ for $CsNiCl_3$, we consider here the
isotropic case $J_\|=J_\bot=1$ for simplicity and since Monte-Carlo simulations
of models with strongly anisotropic parameters require considerably more
computing effort.  For this reason, we also chose a value for $D=-0.2$,
which is small enough for our purposes; the ground states which occur
as a function of H are consistent with those observed
experimentally.\cite{mailA}  (In contrast with the case $D=-1$, for example,
which yields a phase diagram with a completely different
structure.\cite{plumA,mailA})  The present work serves to compliment
and extend the Monte Carlo studies performed at H=0 on this and similar
Hamiltonians,\cite{miya,ohya} as well as one cursory examination of the
phase diagram.\cite{pet}

This study was made in an attempt to accurately estimate the phase-boundary
lines close to the multicritical point at ($T_m,H_m)$.  It
was anticipated to be a numerically challenging problem
since fluctuations involving all three components of {\bf M} are
important in this region of the phase diagram.  The Ferrenberg-Swendsen
histogram method of Monte Carlo simulations offers the possibility of the
precise determination of transition points by the temperature at which
extrema occur in thermodynamic functions for finite-size systems.\cite{ferr}
Relevant components of the staggered susceptibility, defined according
to the components of the order parameter {\bf M} involved in the transition
of interest,\cite{mailA} were used for this purpose.  Simulations were
performed on a lattice of size $12\times12\times12$.  Runs of $1.2\times10^6$
Monte Carlo steps per spin were made, with the initial $2\times10^5$ steps
discarded for thermalization.  For a given value of magnetic field,
single histograms were made at one or more T to ensure that
the maxima in the susceptibility occured close to at least one simulation
temperature.

The results shown in Fig. 1 confirm the general stucture determined by
the phenomenological Landau model \cite{plumC} as well as by a
molecular-field treatment of the Hamiltonian (1).\cite{plumA}  (In the latter
case, two different types of linear and elliptical states (2A, 2B, 3A, and
3B) were distinguished by a relative phase angle, as in the Monte Carlo
studies at H=0.\cite{miya,ohya}  Such distinctions are beyond, and not
relevant to, the goals of the present work.)  The detailed behavior near the
multicritical point at $T_m=0.915(5)$, $H_m=2.62(4)$, however,
is quite different from both mean-field
results.  There is clear indication that both the 1-2 and 1-4 transition
lines approach this point with slopes that are the same in magnitude
as that of the 3-4 spin-flop line.
This is an effect of critical fluctuations and is predicted by scaling
theory.\cite{kawaA}  Less clear is that the 2-3 transition line exhibits
the same predicted tendency.  A remarkable feature of the 1-4 line is
its initial curvature to the left as the field increases.
In the usual case of the bicritical point
associated with unfrustrated antiferromagnets with weak axial anisotropy,
only two critical lines are involved.\cite{fisher}
These also approach this point
asymptotically with slopes that are the same in magnitude as the spin-flop
line and are governed
by the same crossover exponent $\phi$.  In contrast with the present
case, both of these lines have always been found to approach the bicritical
point from the right, forming an umbilicus structure.\cite{shapira}
This behavior has
also been observed in Monte Carlo simulations.\cite{landau}  There appears
to be no argument from scaling theory, however, that relates the {\it signs}
of the initial slopes of critical lines emanating from a multicritical point.
We note that this opposite-slope behavior has not
been observed in $CsNiCl_3$.  This is not surprising since the model
parameters used in here are not relevant for quasi-one-dimensional materials,
a feature which may obscure this unusual effect.

In conclusion, these Monte Carlo results demonstrate that significant
critical-fluctuation effects are associated with this novel multicritcial
point.  While there are strong symmetry arguments to support the conclusion
made in Ref.\ \onlinecite{kawaA} that the 1-2 and 2-3 transitions belong to
the xy and Ising universality classes, respectively, the nature of the 1-4
transition, and that of the multicritical point itself, remain somewhat
unsetteled.  The interpretation made in Ref.\ \onlinecite{kawaA} is that
these transitions are related to the new chiral universality classes
proposed by Kawamura.\cite{kawaB}  (The resulting crossover-exponent value
$\phi \simeq 1.04$ is not inconsistent with the results of Fig. 1.)
Azaria {\it et al} \cite{aza,bhatt,plumD,mailB} have argued
that such transitions exhibit
nonuniversal critical behavior, where a first-order, mean-field tricritical
or $O(4)$ universality can occur.  It is not clear
what type of scaling behavior for the multicritcal point can be expected
in these cases.

\acknowledgements
This work was supported by NSERC of Canada and FCAR du Qu\'ebec.
%==============================================================================
% REFERENCES
%

%=============================================================================
%FIGURES
\begin{figure}
\caption{Phase diagram with ${\bf H}\|{\bf z}\|{\bf c}$ near the
multicritical point at $T_m=0.915(5)$,$H_m=2.62(4)$ as determined
by Monte Carlo simulations (points).
Indicated are the paramagnetic phase 1, linear phase 2 with $M_z$,
elliptical phase 3 with $M_z,M_x$  and spin-flopped
helical phase 4 with $M_x,M_y$.}
\label{fig1}
\end{figure}

%==============================================================================
\end{document}